\documentclass[preprint,aps,onecolumn,floats,floatfix,amsmath,nofootinbib]{revtex4-1}
\usepackage{graphicx,array,dcolumn}
\usepackage{calc,tabularx, epsfig,mathrsfs}
\usepackage{hyperref}

\usepackage{amsmath,verbatim,enumerate}
\usepackage{amssymb}
\usepackage{multirow}
\usepackage{xspace}
\usepackage{slashed}
\allowdisplaybreaks[1]
\newlength{\figurewidth}
\newcommand{\beq}{\begin{equation}}
\newcommand{\eeq}{\end{equation}}
\newcommand{\bea}{\begin{eqnarray}}
\newcommand{\eea}{\end{eqnarray}}
\newcommand{\ba}{\begin{array}}
\newcommand{\ea}{\end{array}}

\newcommand{\pt}{\partial}

\newcommand{\al}{\alpha}
\newcommand{\bt}{\beta}
\newcommand{\g}{\gamma}
\newcommand{\ep}{\epsilon}
\newcommand{\ta}{\theta}
\newcommand{\lam}{\lambda}

\newcommand{\G}{\Gamma}

\newcommand{\de}{\delta}

\newcommand{\OM}{\Omega}

\newcommand{\sg}{\sigma}

\makeatother
\begin{document}
%
\title{Non-local scalar field on deSitter and its infrared behaviour}
\setlength{\figurewidth}{\columnwidth}
%
\author{Gaurav Narain$\,{}^a$}
\email{gaunarain@gmail.com}
\author{Nirmalya Kajuri$\,{}^{b}$}
\email{nirmalyak@gmail.com}
\affiliation{
${}^a$ Department of Space Science, Beihang University, Beijing 100191, China.\\
${}^b$ Chennai Mathematical Institute, Siruseri Kelambakkam 603103.
}
%
%
\begin{abstract}
We investigate free non-local massless and massive scalar field on deSitter (dS) space-time.
We compute the propagator for the non-local scalar field for the corresponding theories 
on flat and deSitter space-times. It is seen that for the non-local theory, the massless limit 
of massive propagator is smooth for both flat and deSitter. Moreover, this limit 
matches exactly with the massless propagator of the non-local scalar field for both 
flat and deSitter space-time. The propagator is seen to respect dS invariance. 
Furthermore, investigations of the non-local Green's function on deSitter 
for large time-like separation shows that the propagator has no infrared divergences. 
The dangerous infrared $\log$-divergent contributions which arise is local massless theories 
are absent in the corresponding non-local version. Lack of infrared divergences 
in the propagator hints at the strong role non-localities may play in the dS infrared physics.  
This study suggest that non-locality can cure IR issues in deSitter. 
\end{abstract}

\maketitle
%
%

\section{Introduction}
\label{intro}

The framework of local quantum field theory (QFT) developed in last century has been 
extremely successful in describing particle physics on flat backgrounds. 
However, problems arise in non-flat backgrounds.  

A particularly interesting case is QFT in de Sitter (dS) space-time. 
This subject is not merely academic - it is strongly 
believed that the early universe went through a phase where it resembled a dS space-time, 
a phase which is commonly known as inflation \cite{Guth:1980zm, Linde:1981mu, Starobinsky:1982ee}. 
Moreover, the experimental data 
of large-scale structures, supernova \cite{Riess:1998cb,Perlmutter:1998np}, 
CMB \cite{Akrami:2018vks,Aghanim:2018eyx} and baryon acoustic oscillations (BAO) 
reveal that currently our Universe is again undergoing a phase of 
accelerated expansion. It is termed as the dark-energy dominated epoch. 
It is therefore extremely important to study 
the behaviour of matter fields on dS space-time.

Quantum matter fields on dS has been a subject of investigation in the last four 
decades \cite{Chernikov:1968zm,Bunch:1978yq,Allen:1985ux,Bros:1994dn} and has seen much 
interest recently \cite{Serreau:2011fu,Bros:2010wa,Marolf:2010zp,Akhmedov:2013vka} 
due to outcome of various experiments
strongly showing that we are in a dS phase. As a results several researchers have tried to 
construct QFT of matter fields on dS. However, there is a problem. 
In the case of free scalar fields it is noticed that the massless limit 
for the free massive propagator is not well defined \cite{Allen:1985wd,Allen:1987tz}, 
in the sense that the massive propagator gets divergent contribution in the massless limit, 
signalling either that the vacuum doesn't respect dS symmetry \cite{Antoniadis:1985pj} or 
one has to `regularise' this divergent piece to get a dS invariant propagator 
\cite{Folacci:1992xc,Kirsten:1993ug}. 
This divergent piece arises due to zero mode contribution \cite{Einhorn:2016fmb} 
in the massless case which renders the limit ill-defined. Gauge-fixing this zero-mode contribution 
leads to an infrared finite Green's function which can be used as an ingredient in the
construction of QFT. 
In interacting theories \cite{Einhorn:2002nu,Rajaraman:2010zx,Hollands:2011we} 
it has been noticed that non-perturbative 
quantum corrections generates a small dynamical mass 
\cite{Gautier:2013aoa,Youssef:2013by} 
which cures the IR behaviour of the effective propagator \cite{Serreau:2011fu}. 
In this paper we take first steps toward investigating whether 
presence of non locality can resolve this IR problem.

There are several motivations for studying non-local field theories in 
de Sitter background. It is known that quantum corrections leading to renormalisation group 
running of couplings with respect to energy can be understood as a 
generation of non-locality due to quantum effects: for example 
if a coupling has running $g(E)$ then it can be thought of as $g(-\Box)$. 
In a sense local theories under quantum corrections lead to non-local theories at low energies. 
Therefore, it makes sense to study the IR behaviour of these theories. 
Recently, nonlocal theories have been 
studied extensively in the context of ultraviolet modification of theories, 
where such non-local modifications may render theory
UV finite or super-renormalizable \cite{Modesto:2011kw,Modesto:2017sdr}. 
Infrared non-locality is mostly studied in the 
context of cosmology to give an alternative explanation to the accelerated expansion 
of cosmological space-time \cite{Deser:2007jk,Capozziello:2008gu,Deser:2013uya,Woodard:2014iga,Elizalde:2011su,Maggiore:2016gpx,Narain:2017twx,Narain:2018hxw}. 
Non-local theories have also been explored in the context of black-hole information 
loss paradox \cite{Kajuri:2017jmy,Kajuri:2018myh}. 
Although non-local infrared modification of QFTs 
has not been explored much, but it is strongly expected that such modifications 
may occur at large distances on dS based on the strong entanglement between 
modes which lie in causally disconnected regions \cite{Maldacena:2012xp,Matsumura:2017swh}. 
It is therefore worth asking whether such non-local modification can lead to 
a well behaved infrared propagator?

In this paper we consider a particular non-local scalar field theory. 
We obtain the Green's function for this theory and show that non-locality 
indeed leads to well behaved IR propagator. This is our main result.

In the next section we introduce the non-local field theory we will study. 
This consist of two parts: flat space-time and deSitter. In the former 
we obtain the Green's function for non-local scalar theory on 
flat space-time (as a trial run), while in later Green's function 
for non-local scalar field is computed on a deSitter background. 
We present our conclusions in the final section.

\section{Non-local scalar field}
\label{NLsca}

In this section we consider a scalar field theory which leads to 
non-local action when one of the scalar gets decoupled from system. 
Consider the following action,
\beq
\label{eq:slocal}
S = \int {\rm d}x \sqrt{-g} \biggl[
\frac{1}{2} (\pt \phi)^2 + \frac{1}{2} (\pt \chi)^2 + \frac{m^2}{2} \phi^2 
- \lam \phi \chi
\biggr] \, ,
\eeq
where $\phi$ and $\chi$ are two scalar fields on curved non-dynamical 
background, $\lam$ is interaction strength and $m$ is mass of scalar $\phi$. 
The equation of motion of two fields give $(-\Box + m^2)\phi - \lam \chi=0$ 
and $-\Box \chi - \lam \phi =0$. Integrating out $\chi$ from the second equation 
of motion yields $\chi = \lam (-\Box)^{-1} \phi$. This when plugged back into 
the action (\ref{eq:slocal}) yields a massive non-local theory for scalar $\phi$. 
This non-local action is given by,
\beq
\label{eq:NLact}
S_{NL} = \int {\rm d}x \sqrt{-g}
\biggl[
\frac{1}{2} (\pt \phi)^2 + \frac{m^2}{2} \phi^2 
-\frac{\lam^2}{2} \phi \frac{1}{-\Box} \phi
\biggr] \, .
\eeq
This non-local action has issues of tachyon. In simple case of 
flat space-time it is noticed that the non-local piece in action reduces 
to $\phi(-p)(-\lam^2/p^2) \phi(p)$ (where $\phi(p)$ is the fourier 
transform of field $\phi(x)$). This piece correspond to something like 
tachyonic mass thereby resulting in issues of unitarity and 
instability of vacuum. However if $\lam^2 \to -\lam^2$ then this 
tachyonic problem gets resolved, although the simple local action 
giving rise to this non-local action will have imaginary coupling. In the following 
we will directly start with the non-local action stated in eq. (\ref{eq:NLact})
with $\lam^2 \to -\lam^2$. This non-local action has no problem of tachyons.
We will study this simple free non-local theory 
on flat and dS space-time. In generic space-time the 
Green's function equation is given by,
\beq
\label{eq:NLgreenFeq}
\biggl(
-\Box + m^2 + \frac{\lam^2}{-\Box}
\biggr) G(x,x^\prime) = -\frac{i \de(x-x^\prime)}{\sqrt{-g}} \, .
\eeq
It should be emphasised that the action for the scalar field $\phi$
given in eq. (\ref{eq:NLact}) doesn't involve dynamical gravity
in the sense that gravity is an external field, and the scalar field 
is living on the fixed background space-time. It shouldn't be confused 
with the scalar fields used to describe the accelerated expansion of Universe.

\subsection{Flat space-time}
\label{flat}

In this section we study this theory on flat space-time. Although our main interest is 
in solving the theory on dS background, it is helpful to first solve it on flat background 
to understand its qualitative features and compare with the dS results.

In flat space-time one can use the momentum space 
representation to write the propagator. Writing 
$G_f(x,x^\prime) = \int {\rm d}p/(2\pi) \tilde{G}(p) 
\exp\{ip(x-x^\prime)\}$, it is seen that 
\beq
\label{eq:GflatNL}
G_f(x,x^\prime)
= -i \int \frac{{\rm d}^dp}{(2\pi)^d}
\biggl(
p^2 + m^2 + \frac{\lam^2}{p^2}
\biggr)^{-1} e^{i p(x-x^\prime)}
\eeq
where the subscript in $G_f$ implies flat space-time. The integrand 
can be written in an alternative manner as following by doing 
partial fraction decomposition. It is given by,
\beq
\label{eq:partFrac}
\bigl(p^2 + m^2 + \lam^2/p^2\bigr)^{-1}
= p^2(p^4 + m^2 p^2 + \lam^2)^{-1}
=A(p^2 + r_{-}^2)^{-1} + B(p^2+r_{+}^2)^{-1} \, ,
\eeq
where the masses $r_\pm^2$ are given by
\beq
\label{eq:rpm}
r_{-}^2 = (m^2 - \sqrt{m^4 - 4\lam^2})/2, 
\hspace{5mm}
r_{+}^2= (m^2 + \sqrt{m^4 - 4\lam^2})/2
\eeq
while the coefficients $A$ and $B$ are 
\beq
\label{eq:ABcoeff}
A = - r_{-}^2/(r_{+}^2 - r_{-}^2) \, ,
\hspace{5mm}
B= r_{+}^2/(r_{+}^2 - r_{-}^2) \,
\eeq
respectively. In the case of massless theory ($m^2\to0$) 
$r_{\pm}^2=\pm i \lam$, $A=-1/2$ and $B=1/2$. 
This partial fraction decompositions is possible if the polynomial  
$(p^4 + m^2 p^2 + \lam^2)$ is not a perfect square,
which is true as long as $m^2 \neq 2\lam$. In the 
situation when $m^2=2\lam$ the coefficients $A$ and $B$ 
diverge as the both $r_{-}^2 = r_{+}^2 = \lam$.
In this case we have 
\beq
\label{eq:sqcase}
\frac{p^2}{(p^2 + \lam)^2} 
= \frac{1}{p^2 + \lam} - \frac{\lam}{(p^2+\lam)^2} \, .
\eeq
It should be noticed that this partial fraction decomposition 
for the case $m^2=2\lam$ can also be obtained from 
eq. (\ref{eq:partFrac}) by writing $m^2 = 2\lam +\rho$ 
and doing a small $\rho$ expansion. This series 
expansion gives,
\begin{align}
\label{eq:smallABrho}
& A = -\frac{1}{2\sqrt{\lam\rho}} \left(\lam - \sqrt{\lam\rho}
+ \frac{3\rho}{8} + \cdots \right) \, ,
\notag \\
& B = \frac{1}{2\sqrt{\lam\rho}} \left(\lam + \sqrt{\lam\rho}
+ \frac{3\rho}{8} + \cdots \right) \, ,
\notag \\
& (p^2 + r_{\pm}^2)^{-1} 
= (p^2 + \lam)^{-1} \mp \sqrt{\lam\rho} (p^2 + \lam)^{-2} + \cdots \, .
\end{align}
It should be noticed that the leading term of $A$ and $B$ is divergent 
as $\rho\to0$. However, when these expansions are plugged back in 
eq. (\ref{eq:partFrac}) it is seen that the divergent term cancels leaving 
behind a finite $\rho\to0$ limit which matches with the RHS in 
eq. (\ref{eq:sqcase}). This observation can be utilised 
in obtaining the Green's function for the case of $m^2=2\lam$
from the Green's function for case $m^2\neq 2\lam$ by taking a limit.

Using inverse Laplace transform one can then rewrite the 
partial fraction form of $\bigl(p^2 + m^2 + \lam^2/p^2\bigr)^{-1}$. 
Then one can express the flat space-time 
Green's function in eq. (\ref{eq:GflatNL}) as follows 
\beq
\label{eq:invLapflat}
G_f(x,x^\prime)
= -i \int \frac{{\rm d}^dp}{(2\pi)^d}
\int_0^{\infty} {\rm d}s 
\bigl[
\cosh(s \bt) - \frac{\al}{\bt} \sinh(s \bt)
\bigr] 
e^{-s(p^2+\al)+i p(x-x^\prime)}
\eeq
where $\al=m^2/2$ and $\bt=\sqrt{m^4-4\lam^2}/2$. 
There are two cases: $m^2>2\lam$ and $m^2<2\lam$.
In the later case $(4\lam^2-m^4=\sg^2>0$ implies 
$r_{-}^2=(m^2 - i \sg)/2$ and $m_{+}^2=(m^2+i \sg)/2$. 
In this case $\bt=i\sg/2$. This means the expression in the square bracket 
in the integrand of eq. (\ref{eq:invLapflat}) becomes 
$\cos(s\sg) -\al/\sg \sin(s\sg)$. In the integrand one 
can smoothly take massless limit to obtain Green's function for non-local 
massless scalar-field. In this massless limit $\al \to0$. 

\begin{figure}[h]
\centerline{
\vspace{0pt}
\centering
\includegraphics[width=4.3in,height=3in]{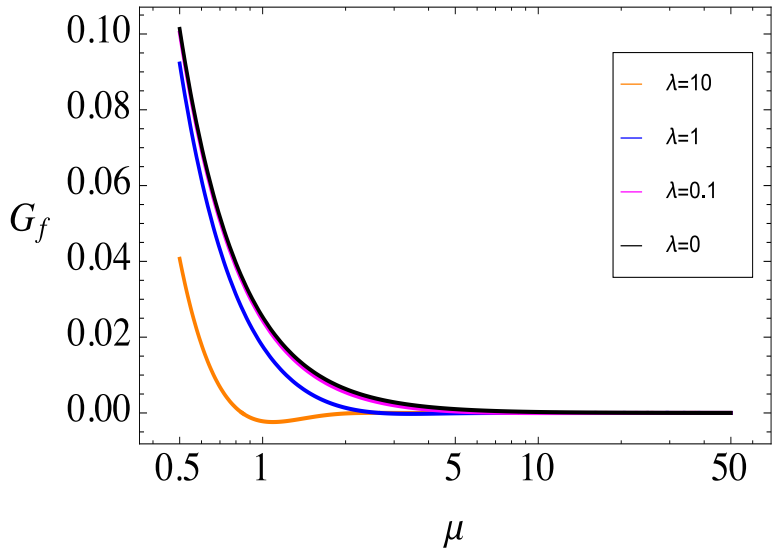}
}
\vspace{-3mm}
\caption[]{
The Green's function of massless non-local scalar field in flat four space-time 
dimensions. The propagator is plotted for various strength of non-locality $\lam$
against the geodesic length $\mu(x,x^\prime)$. 
}
\label{fig:scm0flat}
\end{figure}

\begin{figure}[h]
\centerline{
\vspace{0pt}
\centering
\includegraphics[width=4.3in,height=3in]{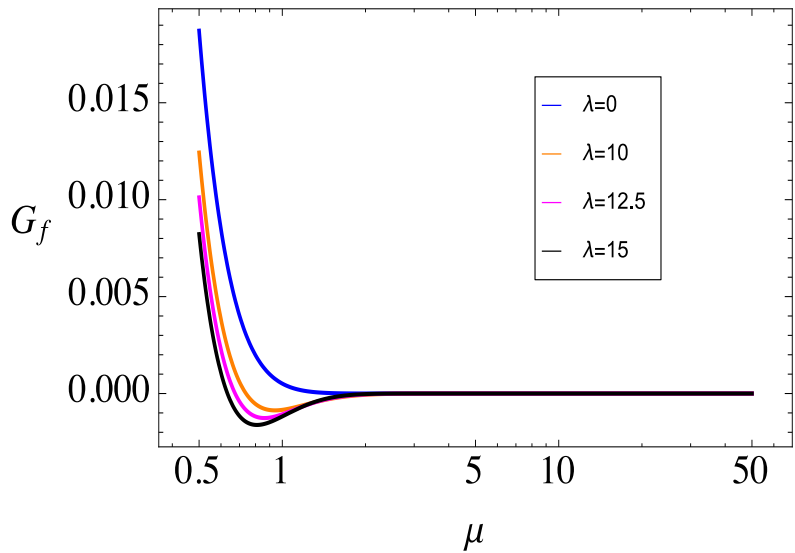}
}
\vspace{-3mm}
\caption[]{
The Green's function of massive scalar field for non-local theory in flat space-time.
The propagator is plotted for various strength of non-locality $\lam$
against the geodesic length $\mu(x,x^\prime)$. The four cases plotted are:
$m^2>2\lam$, $m^2=2\lam$, $m^2<2\lam$ and $\lam=0$ (local). 
}
\label{fig:scmNLflat}
\end{figure}

To evaluate eq. (\ref{eq:invLapflat}) one first perform 
the integration over $p$, then integration over $s$. For $m=0$ case the $s$-integral 
can be performed exactly in closed form in arbitrary dimensions. 
In four dimensions it acquires a simplified form and is given by,
\beq
\label{eq:gflatNL}
G_f(x,x^\prime) =
\frac{\sqrt{i \lam}K_1\left(\sqrt{i \lam}\mu \right)+\sqrt{-i \lambda } 
K_1\left(\sqrt{-i \lam} \mu \right)}{8 \pi ^2 \mu } \, ,
\eeq
where $\mu = \sqrt{(x-x^\prime)^2}$ is the geodesic distance
and $K_1$ is the modified Bessel function of the second kind. 
In Fig. \ref{fig:scm0flat} we plot this massless non-local propagator in flat 
four space-time dimensions for various $\lam$. 
In the limit $\lam\to0$ (when the non-locality is turned off),
the propagator approaches the flat space-time massless propagator 
of local scalar field. 
%
%
In the case when $m\neq0$ things are complicated. Here two possibilities arises 
$m^4>4\lam^2$ and $4\lam^2>m^4$, where both $m$ and $\lam$ are positive. 
In the former case one can take limit $\lam\to0$ (locality limit), while in later case 
one can take the limit $m\to0$ (massless limit). In the locality limit, $r_{-}^2=0$
and $r_{+}^2=m^2$. In this case $A=0$ while $B=1$ ($\al=\bt=m^2/2$). In this case 
one gets the propagator for local massive scalar field which is 
\beq
\label{eq:gfmlam0}
G_f^{\lam=0} = 
(2 \pi )^{-d/2} \left(\frac{m}{\mu}\right)^{(d-2)/2}
K_{\frac{d}{2}-1}(m \mu) \, ,
\eeq
where $K_{d/2-1}$ is the modified Bessel function of the second kind.
It is still possible to perform the above integration 
for $m\neq0$ in closed form. For the massive non-local scalar field in 
arbitrary space-time dimension this Green's function $G_f$ is given by,
\beq
\label{eq:Gfmlam}
G_f^m(x,x^\prime)=
\frac{(2\pi)^{-d/2}\mu^{1-d/2}}{(r_{+}^2-r_{-}^2)} 
\biggl[
r_{+}^{d/2+1} K_{\frac{d}{2}-1} \left(r_{+}\mu\right)
-r_{-}^{d/2+1} K_{\frac{d}{2}-1} \left(r_{-}\mu\right)
\biggr] \, ,
\eeq
where $r_{+}$ and $r_{-}$ are stated before
and $K_{d/2-1}$ is the modified Bessel function of the second kind. In the limit 
$\lam\to0$ this gives smoothly the result stated in eq. (\ref{eq:gfmlam0}).
The $m\to0$ limit of this is smooth and obtains the massless 
non-local propagator for the scalar theory in arbitrary dimension
which agrees with the expression mentioned in eq. (\ref{eq:gflatNL}). 
The expression in arbitrary dimensions is given by,
\bea
\label{eq:Gfd}
&&
G_f(x,x^\prime)=
\frac{i  \left[-\left(i \lam\right)^{\frac{d+2}{4}}
K_{1-\frac{d}{2}}\left(\sqrt{i\lam} \mu \right)
+(-i \lam)^{\frac{d+2}{4}} K_{\frac{d}{2}-1}
\left(\sqrt{-i\lam} \mu \right)\right]}{2 (2\pi)^{d/2}\lambda \mu^{d/2-1}} \, .
\eea
In Fig. \ref{fig:scmNLflat}
we plot the massive Green's function for cases: 
$m^2>2\lam$, $m^2=2\lam$, $m^2<2\lam$ and $\lam=0$ (local). 
The equality case $m^2=2\lam$ is obtained as a limit as has been 
explained before.
%
%
%
\subsection{DeSitter}
\label{ds}

In this section we will investigate the theory eq. (\ref{eq:NLact} on dS background.
The $d$-dimensional deSitter space-time can be identified with the real 
one-sheeted hyperboloid in $(d+1)$ Minkowski space-time $M_{d+1}$:
$X_d= \{x \in \mathbb{R}^{d+1}, x^2= - H^2\}$, where $H$ is the Hubble constant. 
If the two points are denoted by $x$ and $x^\prime$, the length of 
geodesic connecting them is $\mu(x,x^\prime)$. It is useful to introduce 
a quantity $z(x,x^\prime)=\cos^2(H\mu/2)$. For space-like distances 
$\mu^2>0$ results in $0<z<1$, while for time-like separation $\mu^2<0$ 
correspond to $1<z<\infty$. Also by making use of the 
co-ordinates of the embedding space it is noted that the 
\beq 
\label{eq:cosZ}
\cos (H \mu) = Z(x,x^\prime) \, ,
\hspace{5mm}
Z(x,x^\prime)
= \eta_{AB} X^A(x) X^B(x^\prime) \, 
\eeq
is the embedding distance, {\it i.e.} length of chord between the 
point $x$ and $x^\prime$ in the embedding space $\mathbb{R}^{d+1}$. 
The parameter $z(x,x^\prime) = (1 + Z)/2$. 
The nice property of maximally symmetric 
spaces is that Green's function on such space-time become entirely 
a function of $\mu$ (or $z(x,x^\prime)$) instead of being a function of 
both $x$ and $x^\prime$. This allows one to compute Green's function 
exactly by solving linear differential equations. 

In the case of non-local theory, the Green's function equation is given in 
eq. (\ref{eq:NLgreenFeq}). But before we solve for $G(x,x^\prime)$
on dS, we first note the following important identities on curved 
space-time. 
\beq
\label{eq:dsOPiden}
-[\Box^2 - m^2 \Box + \lam^2]^{-1}\Box
= A (-\Box + r_{-}^2)^{-1} + B (-\Box + r_{+}^2)^{-1} \, ,
\eeq
where $A$, $B$, $r_{-}^2$ and $r_{+}^2$ have the same values 
as in flat space-time. This is easy to prove by noticing that 
$-\Box$ can be written in following way
\beq
\label{eq:mBoxIden}
-\Box = A (-\Box + r_+^2) + B (-\Box + r_-^2) \, .
\eeq
This follows from the values of $A$ and $B$ stated in 
eq. (\ref{eq:ABcoeff}) which gives $A r_+^2 + B r_-^2 =0$
and $A+B=1$. Using $[\Box^2 - m^2 \Box + \lam^2]
= (-\Box + r_{-}^2)(-\Box + r_{+}^2)$ one then immediately obtains 
the identity stated in eq. (\ref{eq:dsOPiden}). 
This proof is valid as long as $m^2\neq 2\lam$, in the case 
of equality $m^2=2\lam$ one has follow similar steps 
described previously for flat space-time. In the case of 
equality $m^2=2\lam$, the LHS of eq. (\ref{eq:dsOPiden}) can be 
expressed as,
\beq
\label{eq: opDSequality}
-(\Box - \lam)^{-2}\Box 
= - (\Box - \lam)^{-2} (\Box - \lam + \lam)
= -(\Box - \lam)^{-1} - \lam (\Box - \lam)^{-2} \, .
\eeq
The RHS of eq. (\ref{eq: opDSequality}) can also be obtained from 
RHS of eq. (\ref{eq:dsOPiden}) by writing $m^2 = 2\lam+\rho$ and 
doing a small $\rho$ expansion. Then again the expansion of 
$A$ and $B$ will be given by eq. (\ref{eq:smallABrho}) while we 
have 
\beq
\label{eq:smRhoBox}
(-\Box + r_\pm^2)^{-1} 
= (-\Box + \lam)^{-1} \mp \sqrt{\lam\rho} (-\Box + \lam)^{-2} + \cdots \, .
\eeq
On plugging these expansions back in eq. (\ref{eq:dsOPiden}) 
it is noticed that divergent piece cancels (as in flat space-time) 
leaving behind a finite $\rho\to0$ limit, which matches the 
RHS of eq. (\ref{eq: opDSequality}). This observation will be useful later
in computing the Green's function on the dS for the case of 
$m^2=2\lam$.

The identity in eq. (\ref{eq:dsOPiden}) is valid for $\lam\neq0$. In the case 
when $\lam=0$ the operator on LHS reduces to just $\Box + m^2$
which is the operator for the local massive scalar field. 
It is the identify in eq. (\ref{eq:dsOPiden}) which allows us to compute the 
Green's function of the non-local theory in deSitter space-time. This means 
that the full Green's function for non-local theory $G_{NL}(x,x^\prime)$
is a sum of two Green's function 
\beq
\label{eq:GNLsum}
G_{NL}(x,x^\prime) = A G_1(x,x^\prime) + B G_2(x,x^\prime) \, ,
\eeq
where 
\beq
\label{eq:greenG1G2}
(-\Box + r_{-}^2)G_1(x,x^\prime)=0 \, ,
\hspace{5mm} 
(-\Box + r_{+}^2)G_2(x,x^\prime)=0 \, .
\eeq
These two Green's function 
can be easily computed using the standard methods of computing 
Green's function for massive scalar field on deSitter space-time.
For example, in the case of local massive scalar field we have 
$(-\Box+m^2)G(x,x^\prime)=0$ as Green's function equation. 
In this case by exploiting the knowledge 
that $G(x,x^\prime)=G(z(x,x^\prime))$, the operator when acting on $G$
acquires the following hyper-geometric form
\beq
\label{eq:GmOP}
z(1-z)G^{\prime\prime}
+\left[c-(a+b+1)z\right] G^\prime - ab G=0 \, ,
\eeq
where $G^\prime = {\rm d} G(z)/{\rm d}z$, 
$a=\bigl[d-1+\bigl((d-1)^2 - 4m^2/H^2 \bigr)^{1/2}\bigr]/2$, 
$b=\bigl[d-1-\bigl((d-1)^2 - 4m^2/H^2 \bigr)^{1/2}\bigr]/2$
and $c=d/2$. This is a Hyper-geometric differential equation 
of second order and has two linearly independent solution:
${}_2F_1(a,b,c;z)$ and ${}_2F_1(a,b,c;1-z)$ \cite{Allen:1985ux,Allen:1985wd,Allen:1987tz}. 
The former has a singularity at $z=1$ (which corresponds to short distance 
$\mu=0$) while the later is singular at $z=0$ (corresponding to 
antipodal separation). By requiring the short distance singularity of 
dS propagator to match with the singular behaviour of the 
flat space-time propagator, and being regular at $z=0$, one concludes 
that $G(z) = q \times {}_2F_1 (a,b,c;z)$ (where $q$ is to be fixed by requiring that 
the strength of singularity of dS propagator to match with the strength of 
singularity in flat space-time). The coefficient $q$ is given by,
\beq
\label{eq:Qval}
q = \frac{H^{d-2} \G(a) \G(b)}{(4\pi)^{d/2} \G(c)} \, .
\eeq
Using this one can write the Green's function $G_{NL}$ for 
non-local scalar on dS to be
\beq
\label{eq:GNLmdSform}
G_{NL}(z) = A q \, {}_2F_1(a_1,b_1,c_1;z)
+ B \bar{q} \,  {}_2F_1(a_2,b_2,c_2;z) \, ,
\eeq
where the coefficients $q$ and $\bar{q}$ are 
obtained from eq. (\ref{eq:Qval}) by making transformation 
$m^2 \to r_{-}^2$ and $m^2 \to r_{+}^2$ respectively, while 
\begin{align}
\label{eq:abc12}
& a_1 = \frac{1}{2}\bigl[d-1+\sqrt{(d-1)^2 - \frac{4 r_{-}^2}{H^2}} \bigr] \, ,
& b_1 =\frac{1}{2}\bigl[d-1-\sqrt{(d-1)^2 - \frac{4 r_{-}^2}{H^2}} \bigr] \, ,
& c_1 = d/2 \, , 
\notag
\\
& a_2 = \frac{1}{2} \bigl[d-1+\sqrt{(d-1)^2 - \frac{4 r_{+}^2}{H^2}} \bigr]\, ,
& b_2 = \frac{1}{2} \bigl[d-1-\sqrt{(d-1)^2 - \frac{4 r_{+}^2}{H^2}} \bigr] \, ,
& c_2 = d/2 \, . 
\end{align}
The values of $q$ and $\bar{q}$ so obtained can be plugged 
back in eq. (\ref{eq:GNLmdSform}) to obtain the full 
non-local Green's function on dS for space-like separation ($0<z<1$).
This is the propagator for arbitrary mass $m$, $\lam$ and $d$. 

\begin{figure}[h]
\centerline{
\vspace{0pt}
\centering
\includegraphics[width=4.3in,height=3in]{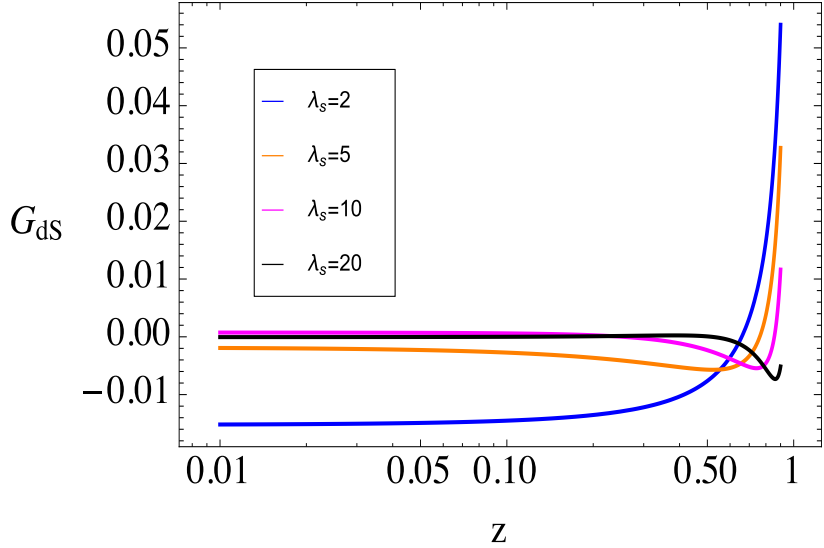}
}
\vspace{-3mm}
\caption[]{
The Green's function of massless scalar field for non-local theory in deSitter space-time.
The propagator is plotted for various strength of non-locality $\lam_s = \lam/H^2$
against $z(x,x^\prime)$. 
}
\label{fig:scm0NLdS}
\end{figure}

\begin{figure}[h]
\centerline{
\vspace{0pt}
\centering
\includegraphics[width=4.3in,height=3in]{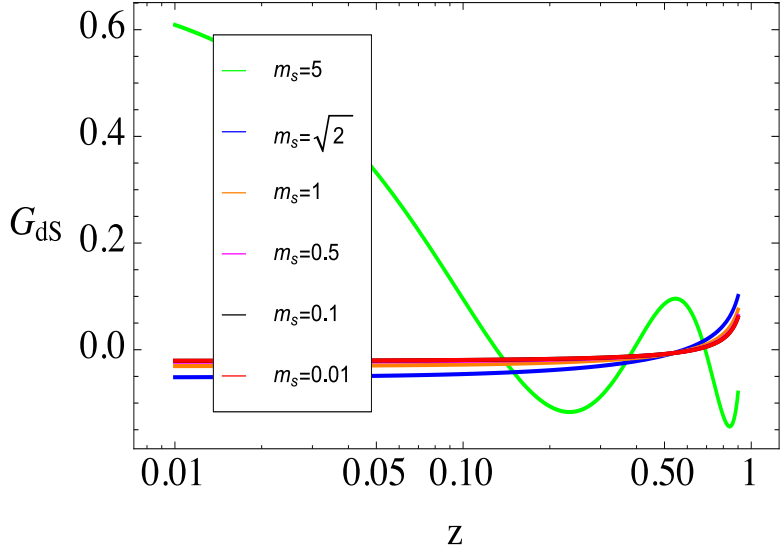}
}
\vspace{-3mm}
\caption[]{
The Green's function of non-local scalar field theory in deSitter space-time.
Here we slowly take the $m\to0$ limit. We plot the three cases: $m^2>2\lam$,
$m^2=2\lam$ and $m^2<2\lam$. In the last case we plot the propagator 
for decreasing values of $m$ to indicate the smooth massless limit. 
The propagator is plotted for fixed $\lam$ and various values of decreasing $m$.
}
\label{fig:GmNLmto0}
\end{figure}

The beautiful thing about the presence of non-locality (however small)
is that it allows us to take a smooth massless limit on deSitter,
which is not possible in case of local scalar field theory 
on deSitter. This is the main and most important result of this paper. 
Moreover, the Green's function for non-local massless scalar field computed 
directly (by starting from a non-local massless theory directly), agrees 
exactly with the case of massive non-local Green's function in limit 
$m\to0$. This beautiful smooth limit is due to the lack of zero modes in the case of 
non-local scalar field theory, which is not so in local case. 
This massless limit is given by,
\bea
\label{eq:Gds4Dm0}
\left. G_{NL} \right|_{m\to0} = \frac{H^{d-2}}{2 (4\pi)^{d/2}} \left[
\G(a_1)\G(b_1) {}_2 \tilde{F}_1 (a_1,b_1,c_1;z)
+ \G(a_2)\G(b_2) {}_2 \tilde{F}_1 (a_2,b_2,c_2;z)
\right] \, ,
\eea
where the parameters $a_1$, $b_1$, $c_1$, $a_2$, $b_2$ 
and $c_2$ here are determined from eq. (\ref{eq:abc12})
in the limit $m\to0$. 
In Figure \ref{fig:scm0NLdS} we plot this Green's function 
for various values of $\lam_s$. 
%
%
For the case of non-zero mass, the propagator has one more parameter, however 
structure remains same. Here we have three cases
(as in flat space-time): $m^2<2\lam$, $m^2=2\lam$ and $m^2>2\lam$. 
In each case the propagator is real for space-like separation.
It is worthwhile to plot the propagator for fixed value of $\lam_s$ and 
for decreasing mass. It is seen that as $m\to0$, the massive 
propagator smoothly approaches the massless non-local propagator. 
In figure. \ref{fig:GmNLmto0} we plot this scenario. 
%
%
The case of equality is interesting as then the propagator depends only on 
one parameter. Physically it implies that the two length scales are 
comparable. In this case the non-locality scale is coupled with the mass, so 
$m\to0$ implies $\lam\to0$ simultaneously.  
In this equality case, as has been discussed before, 
we have $r_{-}^2 = r_{+}^2 = m^2/2$. 
This implies that $a_1=a_2$, $b_1=b_2$, while $A$ and $B$ diverges. 
This divergence in $A$ and $B$ has been discussed before, 
and it is not a problematic thing as it cancels off. 
As before we can define $m^2 = 2\lam + \rho$ and compute the propagator.
Expanding the propagator in powers of $\rho$ clearly shows the absence of 
any divergence which is expected as shown previously, the way 
eq. (\ref{eq:dsOPiden}) reduces to eq. (\ref{eq: opDSequality}) 
in the limit $\rho\to0$. The Green's function can then be obtained 
by taking the limit $\rho\to0$. The Green's function obtained in this limit is well-defined 
and analytic. In figure \ref{fig:GmNLmto0} the case of $m_s = m/H = \sqrt{2}$
refers to this scenario. 

To obtain the propagator for time-like separation (or Feynman propagator)
we consider two cases: $G_{NL}(z+i\ep)$ and $G_{NL}(z-i\ep)$ (as for 
$z>1$ the hypergeometric function has branch-cut on the real axis) 
\cite{Allen:1985ux,Allen:1985wd,Allen:1987tz}. 
The Feynman propagator $G_{NL}^F(z) = \lim_{\ep\to0}G_{NL}(z+i\ep)$, while 
the symmetric propagator is $G_{NL}^S(z) = 
\lim_{\ep\to0}\{G_{NL}(z+i\ep) + G_{NL}(z-i\ep)\}$. 

\subsection{Large time behaviour}
\label{largeT}

It is important to check how the propagator in non-local theory 
behaves at large time. To investigate this one has to write the 
propagator in Lorentzian signature. For this one can first write the 
dS metric in either flat or static co-ordinates system. We consider 
static co-ordinate system:
\beq
\label{eq:staticC}
{\rm d}s^2 = -(1-H^2 r^2) {\rm d}t^2 
+ \frac{{\rm d}r^2}{(1-H^2 r^2)}
+ r^2 {\rm d}\OM^2 \, .
\eeq
The embedding map between the 5D bulk space 
$X^\al$ ($\al=0,\cdots,4$) and the static patch 
$x^\mu= (t,r,\ta,\phi)$ is:
\begin{align}
\label{eq:patchTR}
& X^0 = \sqrt{1- H^2 r^2} \sinh(H t) \, ,
&& X^1 = r \sin \ta \cos \phi \, , 
& X^2 = r \sin\ta \sin\phi \, , 
\notag \\
& X^3 = r \cos\ta \, ,
&& X^4 = \sqrt{1-H^2 r^2} \cosh(H t) \, ,
\end{align}
where $1\leq H r \leq 1$, $0\leq\ta\leq\pi$ and $0\leq\phi\leq2\pi$. 
These transformation then imply that the parameter $z(x,x^\prime)$
can be computed using eq. (\ref{eq:cosZ}) which is given by
\beq
\label{eq:zstatic}
2 z -1 = \sqrt{(1 - H^2 r^2)(1 - H^2 r^{\prime 2})} \cosh(H(t-t^\prime))
+ H^2 r r^\prime \cos \OM \, ,
\eeq
where 
\beq
\label{eq:cosOM}
\cos \OM \equiv \cos \ta \cos \ta^\prime
+ \sin \ta \sin\ta^\prime \cos(\phi-\phi^\prime) \, .
\eeq
Then to study the large time behaviour of the Green's function 
$G(x,x^\prime)$ we just fix one co-ordinate and take the other to
large distance. To achieve this we choose the following configuration 
\beq
\label{eq:largeTconfig}
x: \{ t= T, r= r, \ta=0, \phi=0 \} \, ,
\hspace{5mm}
x^\prime : \{t^\prime=0, r^\prime =r, \ta^\prime=0, \phi^\prime =0 \} \, .
\eeq
This gives for time-like separation (where one has to take 
in to account the proper $i\ep$-prescription)
\beq
\label{eq:Tz}
z(x,x^\prime) = \frac{1+Z}{2} 
= \cosh^2\left(\frac{HT}{2}\right)
- H^2 r^2 \sinh^2\left(\frac{HT}{2}\right) + i \ep
\approx \frac{e^{H T}}{4} (1 - H^2 r^2) + i \ep \, .
\eeq
Using this one can now study the large $T$ behaviour of the 
Green's function by plugging $z(x,x^\prime)$ in to the expression for 
Green's function $G_{NL}(z)$. Furthermore, if $r=0$ then $z$ only 
gets contribution from $T$. In this sense for time-like separation,
the large $T$ corresponds to large $z(x,x^\prime)$. Our Green's function 
is a linear combination of two hyper-geometric functions as indicated in 
eq. (\ref{eq:GNLmdSform}) where $q$ and $\bar{q}$ are constants 
depending on parameters of the theory. The large-$z$ expansion 
of hypergeometric functions are well known in literature. For example 
for ${}_2F_1(a,b,c;z)$ in $z\to\infty$ gives 
\begin{align}
\label{eq:largezF}
{}_2F_1(a,b,c;z) = &\frac{\G(c) \G(b-a)}{\G(b) \G(c-a)} (-z)^{-a} 
{}_2 F_1(a,a-c+1,a-b+1;z^{-1}) 
\notag \\
&
+ \frac{\G(c) \G(a-b)}{\G(a) \G(c-b)} (-z)^{-b} 
{}_2 F_1(b,b-c+1,b-a+1;z^{-1}) \, ,
\end{align}
where $\lvert \arg(1-z) \rvert <\pi$. This is true when $(a-b)$ is not an integer. 
For massless local scalar theory this condition is not satisfied 
as $(a-b)$ is an integer in any space-time dimensions. Then the 
asymptotic expansion of hyper-geometric functions gets an 
additional factor of $\log z$. This $\log z$ factor is the source of 
IR divergence when $z\to\infty$. In case when non-locality is 
present $(a-b)$ is never integer for any space-time dimensions.
As a result $\log z$ factor doesn't arise in asymptotic expansion of 
hyper-geometric functions thereby implying absence of dangerous 
IR $\log$-divergence. 

This identity can be utilised to express the large-$T$ behaviour of 
the non-local Green's function. This will imply that for $z\to\infty$
\begin{align}
\label{eq:largeGNL}
G_{NL}(z) = &\frac{q \G(c_1) \G(b_1-a_1)}{\G(b_1) \G(c_1-a_1)} (-z)^{-a_1} 
{}_2 F_1(a_1,a_1-c_1+1,a_1-b_1+1;z^{-1}) 
\notag \\
& + \frac{q \G(c_1) \G(a_1-b_1)}{\G(a_1) \G(c_1-b_1)} (-z)^{-b_1} 
{}_2 F_1(b_1,b_1-c_1+1,b_1-a_1+1;z^{-1}) 
\notag \\
&+ \frac{\bar{q} \G(c_2) \G(b_2-a_2)}{\G(b_2) \G(c_2-a_2)} (-z)^{-a_2} 
{}_2 F_1(a_2,a_2-c_2+1,a_2-b_2+1;z^{-1}) 
\notag \\
&
+ \frac{\bar{q} \G(c_2) \G(a_2-b_2)}{\G(a_2) \G(c_2-b_2)} (-z)^{-b_2} 
{}_2 F_1(b_2,b_2-c_2+1,b_2-a_2+1;z^{-1}) \, .
\end{align}
Plugging eq. (\ref{eq:Tz}) in eq. (\ref{eq:largeGNL}) we get the 
behaviour of the non-local Green's function at large time-like 
separations. This is given by,
\begin{align}
\label{eq:GNL_LT}
&
G_{NL} \rvert_{z\to\infty} = 
\frac{q \G(c_1) \G(b_1-a_1) (-\g)^{-a_1}}{\G(b_1) \G(c_1-a_1)} e^{-a_1 HT} 
+ \frac{q \G(c_1) \G(a_1-b_1)(-\g)^{-b_1}}{\G(a_1) \G(c_1-b_1)} e^{-b_1 HT}
\notag \\
&
+\frac{\bar{q} \G(c_2) \G(b_2-a_2) (-\g)^{-a_2}}{\G(b_2) \G(c_2-a_2)} e^{-a_2 HT} 
+\frac{\bar{q} \G(c_2) \G(a_2-b_2) (-\g)^{-b_2}}{\G(a_2) \G(c_2-b_2)} e^{-b_2 HT} \, ,
\end{align}
where $\g = (1-H^2 r^2)/4$. 
As $T\to\infty$ each of these terms get exponentially suppressed resulting in 
IR well-behaved Green's function even for massless theory. This is possible 
due to the presence of non-locality. This implies that the Green's function on dS in the 
presence of non-locality respect cluster-decomposition theorem as correlation 
exponentially decays at large time-like separations.

\section{Conclusions}
\label{conc}

In this letter we considered non-local scalar field theory on flat and de Sitter space-time.
We computed the propagator of scalar field in either case, for both massless and massive 
theories. It is seen that the limit $m\to0$ is smooth in presence of non-locality on dS.
Moreover, the $m\to0$ limit of propagator of massive non-local scalar field 
matches exactly with the massless non-local propagator for both flat and dS space-time. 
This offers an interesting solution to the long-standing problem of infrared divergence 
for scalar field theories on dS. It is seen that no divergence arises in the limit $m\to0$
in the presence of non-locality however small. Furthermore, it is seen that the 
correlation of two fields decays exponentially at large time-like separations, 
and the presence of non-locality doesn't give rise to dangerous 
$\log z$ divergent in massless theories which is the case in local theories. 
In this way the presence of non-locality leads to infrared 
divergence free Green's function. It shows that the 
Green's function of the non-local theory obeys cluster decomposition theorem. 
These are the two most important results of the paper. 
It shows that non-locality may hold the key to cure the IR issues of field theory 
in dS space-time. 

This opens up several future directions. There has been no systematic study of 
low energy non-local modifications of field theories in de Sitter space-time. 
Our result suggests that this subject merits further investigation. 
It is interesting to ask whether this result holds 
for a larger class of non-local field theories. Furthermore, it will be worth 
investigating interacting theories in the presence of non-locality 
and study low energy effective action of such theories which is expected 
to get modified due to IR non-locality at large distances.

\bigskip
\centerline{\bf Acknowledgements} 

GN will like to thank Alok Laddha for useful discussions. 
NK is supported by the SERB National Postdoctoral Fellowship. 
GN is supported by ``Zhuoyue" (Distinguished) Fellowship (ZYBH2018-03).

\appendix




\begin{thebibliography}{99} 
%
%
\bibitem{Guth:1980zm} 
  A.~H.~Guth,
  Phys.\ Rev.\ D {\bf 23}, 347 (1981)
  [Adv.\ Ser.\ Astrophys.\ Cosmol.\  {\bf 3}, 139 (1987)].
  doi:10.1103/PhysRevD.23.347


\bibitem{Linde:1981mu} 
  A.~D.~Linde,
  Phys.\ Lett.\  {\bf 108B}, 389 (1982)
  [Adv.\ Ser.\ Astrophys.\ Cosmol.\  {\bf 3}, 149 (1987)].
  doi:10.1016/0370-2693(82)91219-9


\bibitem{Starobinsky:1982ee} 
  A.~A.~Starobinsky,
  Phys.\ Lett.\  {\bf 117B}, 175 (1982).
  doi:10.1016/0370-2693(82)90541-X


\bibitem{Riess:1998cb} 
  A.~G.~Riess {\it et al.} [Supernova Search Team],
  Astron.\ J.\  {\bf 116}, 1009 (1998)
  doi:10.1086/300499
  [astro-ph/9805201].


\bibitem{Perlmutter:1998np} 
  S.~Perlmutter {\it et al.} [Supernova Cosmology Project Collaboration],
  Astrophys.\ J.\  {\bf 517}, 565 (1999)
  doi:10.1086/307221
  [astro-ph/9812133].


\bibitem{Akrami:2018vks} 
  Y.~Akrami {\it et al.} [Planck Collaboration],
  arXiv:1807.06205 [astro-ph.CO].


\bibitem{Aghanim:2018eyx} 
  N.~Aghanim {\it et al.} [Planck Collaboration],
  arXiv:1807.06209 [astro-ph.CO].


\bibitem{Chernikov:1968zm} 
  N.~A.~Chernikov and E.~A.~Tagirov,
  Ann.\ Inst.\ H.\ Poincare Phys.\ Theor.\ A {\bf 9}, 109 (1968).


\bibitem{Bunch:1978yq} 
  T.~S.~Bunch and P.~C.~W.~Davies,
  Proc.\ Roy.\ Soc.\ Lond.\ A {\bf 360}, 117 (1978).
  doi:10.1098/rspa.1978.0060


\bibitem{Allen:1985ux} 
  B.~Allen,
  Phys.\ Rev.\ D {\bf 32}, 3136 (1985).
  doi:10.1103/PhysRevD.32.3136


\bibitem{Bros:1994dn} 
  J.~Bros, U.~Moschella and J.~P.~Gazeau,
  Phys.\ Rev.\ Lett.\  {\bf 73}, 1746 (1994).
  doi:10.1103/PhysRevLett.73.1746


\bibitem{Serreau:2011fu} 
  J.~Serreau,
  Phys.\ Rev.\ Lett.\  {\bf 107}, 191103 (2011)
  doi:10.1103/PhysRevLett.107.191103
  [arXiv:1105.4539 [hep-th]].


\bibitem{Bros:2010wa} 
  J.~Bros, H.~Epstein and U.~Moschella,
  Lett.\ Math.\ Phys.\  {\bf 93}, 203 (2010)
  doi:10.1007/s11005-010-0406-4
  [arXiv:1003.1396 [hep-th]].


\bibitem{Marolf:2010zp} 
  D.~Marolf and I.~A.~Morrison,
  Phys.\ Rev.\ D {\bf 82}, 105032 (2010)
  doi:10.1103/PhysRevD.82.105032
  [arXiv:1006.0035 [gr-qc]].


\bibitem{Akhmedov:2013vka} 
  E.~T.~Akhmedov,
  Int.\ J.\ Mod.\ Phys.\ D {\bf 23}, 1430001 (2014)
  doi:10.1142/S0218271814300018
  [arXiv:1309.2557 [hep-th]].


\bibitem{Allen:1985wd} 
  B.~Allen and T.~Jacobson,
  Commun.\ Math.\ Phys.\  {\bf 103}, 669 (1986).
  doi:10.1007/BF01211169


\bibitem{Allen:1987tz} 
  B.~Allen and A.~Folacci,
  Phys.\ Rev.\ D {\bf 35}, 3771 (1987).
  doi:10.1103/PhysRevD.35.3771


\bibitem{Antoniadis:1985pj} 
  I.~Antoniadis, J.~Iliopoulos and T.~N.~Tomaras,
  Phys.\ Rev.\ Lett.\  {\bf 56}, 1319 (1986).
  doi:10.1103/PhysRevLett.56.1319


\bibitem{Folacci:1992xc} 
  A.~Folacci,
  Phys.\ Rev.\ D {\bf 46}, 2553 (1992)
  doi:10.1103/PhysRevD.46.2553
  [arXiv:0911.2064 [gr-qc]].


\bibitem{Kirsten:1993ug} 
  K.~Kirsten and J.~Garriga,
  Phys.\ Rev.\ D {\bf 48}, 567 (1993)
  doi:10.1103/PhysRevD.48.567
  [gr-qc/9305013].


\bibitem{Einhorn:2016fmb} 
  M.~B.~Einhorn and D.~R.~T.~Jones,
  JHEP {\bf 1703}, 144 (2017)
  doi:10.1007/JHEP03(2017)144
  [arXiv:1606.02268 [hep-th]].


\bibitem{Einhorn:2002nu} 
  M.~B.~Einhorn and F.~Larsen,
  Phys.\ Rev.\ D {\bf 67}, 024001 (2003)
  doi:10.1103/PhysRevD.67.024001
  [hep-th/0209159].


\bibitem{Rajaraman:2010zx} 
  A.~Rajaraman, J.~Kumar and L.~Leblond,
  Phys.\ Rev.\ D {\bf 82}, 023525 (2010)
  doi:10.1103/PhysRevD.82.023525
  [arXiv:1002.4214 [hep-th]].


\bibitem{Hollands:2011we} 
  S.~Hollands,
  Annales Henri Poincare {\bf 13}, 1039 (2012)
  doi:10.1007/s00023-011-0140-1
  [arXiv:1105.1996 [gr-qc]].


\bibitem{Gautier:2013aoa} 
  F.~Gautier and J.~Serreau,
  Phys.\ Lett.\ B {\bf 727}, 541 (2013)
  doi:10.1016/j.physletb.2013.10.072
  [arXiv:1305.5705 [hep-th]].


\bibitem{Youssef:2013by} 
  A.~Youssef and D.~Kreimer,
  Phys.\ Rev.\ D {\bf 89}, 124021 (2014)
  doi:10.1103/PhysRevD.89.124021
  [arXiv:1301.3205 [gr-qc]].


\bibitem{Modesto:2011kw} 
  L.~Modesto,
  Phys.\ Rev.\ D {\bf 86}, 044005 (2012)
  doi:10.1103/PhysRevD.86.044005
  [arXiv:1107.2403 [hep-th]].


\bibitem{Modesto:2017sdr} 
  L.~Modesto and L.~Rachwa?,
  Int.\ J.\ Mod.\ Phys.\ D {\bf 26}, no. 11, 1730020 (2017).
  doi:10.1142/S0218271817300208


\bibitem{Deser:2007jk} 
  S.~Deser and R.~P.~Woodard,
  Phys.\ Rev.\ Lett.\  {\bf 99}, 111301 (2007)
  doi:10.1103/PhysRevLett.99.111301
  [arXiv:0706.2151 [astro-ph]].


\bibitem{Capozziello:2008gu} 
  S.~Capozziello, E.~Elizalde, S.~Nojiri and S.~D.~Odintsov,
  Phys.\ Lett.\ B {\bf 671}, 193 (2009)
  doi:10.1016/j.physletb.2008.11.060
  [arXiv:0809.1535 [hep-th]].


\bibitem{Deser:2013uya} 
  S.~Deser and R.~P.~Woodard,
  JCAP {\bf 1311}, 036 (2013)
  doi:10.1088/1475-7516/2013/11/036
  [arXiv:1307.6639 [astro-ph.CO]].


\bibitem{Woodard:2014iga} 
  R.~P.~Woodard,
  Found.\ Phys.\  {\bf 44}, 213 (2014)
  doi:10.1007/s10701-014-9780-6
  [arXiv:1401.0254 [astro-ph.CO]].


\bibitem{Elizalde:2011su} 
  E.~Elizalde, E.~O.~Pozdeeva and S.~Y.~Vernov,
  Phys.\ Rev.\ D {\bf 85}, 044002 (2012)
  doi:10.1103/PhysRevD.85.044002
  [arXiv:1110.5806 [astro-ph.CO]].


\bibitem{Maggiore:2016gpx} 
  M.~Maggiore,
  Fundam.\ Theor.\ Phys.\  {\bf 187}, 221 (2017)
  doi:10.1007/978-3-319-51700-116
  [arXiv:1606.08784 [hep-th]].


\bibitem{Narain:2017twx} 
  G.~Narain and T.~Li,
  Phys.\ Rev.\ D {\bf 97}, no. 8, 083523 (2018)
  doi:10.1103/PhysRevD.97.083523
  [arXiv:1712.09054 [hep-th]].


\bibitem{Narain:2018hxw} 
  G.~Narain and T.~Li,
  Universe {\bf 4}, no. 8, 82 (2018)
  doi:10.3390/universe4080082
  [arXiv:1807.10028 [hep-th]].


\bibitem{Kajuri:2017jmy} 
  N.~Kajuri,
  Phys.\ Rev.\ D {\bf 95}, no. 10, 101701 (2017)
  doi:10.1103/PhysRevD.95.101701
  [arXiv:1704.03793 [gr-qc]].


\bibitem{Kajuri:2018myh} 
  N.~Kajuri and D.~Kothawala,
  arXiv:1806.10345 [gr-qc].


\bibitem{Maldacena:2012xp} 
  J.~Maldacena and G.~L.~Pimentel,
  JHEP {\bf 1302}, 038 (2013)
  doi:10.1007/JHEP02(2013)038
  [arXiv:1210.7244 [hep-th]].


\bibitem{Matsumura:2017swh} 
  A.~Matsumura and Y.~Nambu,
  Phys.\ Rev.\ D {\bf 98}, no. 2, 025004 (2018)
  doi:10.1103/PhysRevD.98.025004
  [arXiv:1707.08414 [gr-qc]].
%
%
\end{thebibliography}
\end{document}